\documentclass[5p,twocolumn,number,sort&compress]{elsarticle}
\usepackage{dcolumn}
\usepackage{amsmath,amssymb,amsthm}
\usepackage{graphicx}
\usepackage{xcolor}
\usepackage{ascmac}
\usepackage[breaklinks=true,linktocpage=true]{hyperref} 
\usepackage{mathrsfs}

\newcommand{\trr}{\triangleright}
\newcommand{\tf}{\tfrac}

\newcommand{\vevs}[1]{\langle #1 \rangle}

\newcommand{\link}{\,{\rm Link}\,}

\newcommand{\er}[1]{Eq.~\eqref{#1}}
\newcommand{\ers}[1]{Eqs.~\eqref{#1}}

\newcommand{\bb}{\mathbb}

\newcommand{\hph}{\hphantom}

\newcommand{\der}{\partial}

\renewcommand{\(}{\left(}
\renewcommand{\)}{\right)}

\newcommand{\wed}{\wedge}

\begin{document}
\allowdisplaybreaks[4]

\title{Higher-form symmetries and 3-group in axion electrodynamics}

\author[1,2,3]{Yoshimasa Hidaka}
\ead{hidaka@post.kek.jp}
\author[4]{Muneto Nitta}
\ead{nitta@phys-h.keio.ac.jp}
\author[1,4]{Ryo Yokokura}
\ead{ryokokur@post.kek.jp}
\address[1]{KEK Theory Center, Tsukuba 305-0801, Japan}
\address[2]{Graduate University for Advanced Studies (Sokendai), 
Tsukuba 305-0801, Japan}
\address[3]{RIKEN iTHEMS, RIKEN, Wako 351-0198, Japan}
\address[4]{Department of Physics \& Research and Education Center for
Natural Sciences, Keio University, 
\par
Hiyoshi 4-1-1, Yokohama, Kanagawa
223-8521, Japan}
\tnotetext[t0]{KEK-TH-2232}

\date{\today}

\begin{abstract}
We study higher-form symmetries in a low-energy effective 
theory of a massless axion coupled with a photon 
in $(3+1)$ dimensions.
It is shown that the higher-form symmetries of this system 
are accompanied by a semistrict 3-group (2-crossed module) 
structure, which can be found by 
the correlation functions of symmetry generators of
 the higher-form symmetries.
We argue that 
the Witten effect and anomalous Hall effect in 
the axion electrodynamics
can be described in terms of 3-group transformations.
\end{abstract}
\maketitle
 \setcounter{footnote}{0}%
\def\thefootnote{$*$\arabic{footnote}}%
   \def\@makefnmark{\hbox
       to\z@{$\m@th^{\@thefnmark}$\hss}}%
\section{Introduction}

Axion electrodynamics has been extensively 
studied in many contexts of modern physics
from particle 
physics~\cite{Peccei:1977hh,Wilczek:1977pj,Weinberg:1977ma,Dine:1981rt,Zhitnitsky:1980tq,Kim:1979if,Shifman:1979if} 
to condensed matter physics~\cite{Wilczek:1987mv,Essin:2008rq,Qi:2008ew}
(see, e.g., Refs.~\cite{Kim:1986ax,Dine:2000cj,Peccei:2006as,Hasan:2010xy,Kawasaki:2013ae} 
as a review).
One of the characteristic features of the axion electrodynamics 
is a topological coupling between
a photon and an axion 
through the chiral anomaly of Dirac fermions coupled with 
them.
Such a coupling modifies  the electric Gauss law and 
the Maxwell-Amp\`ere law. 
Another feature is that
there can be
spatially or temporally extended solitonic objects 
such as a Dirac monopole, an axionic string (dislocation) and 
a domain wall~\cite{Davis:1986xc,Qi:2008ew}.
Due to the modified Maxwell equations, 
the solitonic objects exhibit rich physical 
phenomena such as 
the Witten effect~\cite{Witten:1979ey,Fischler:1983sc,Sikivie:1984yz,Kogan:1992bq} 
for the Gauss law
and the anomalous Hall effect~\cite{Wilczek:1987mv,Essin:2008rq,Qi:2008ew,Teo:2010zb,Zyuzin:2012tv,Wang:2012bgb,Ferrer:2015iop,Yamamoto:2015maz,Qiu:2016hzd,Bednik:2016phb,Ferrer:2016toh}
around an axionic string~\cite{Qi:2008ew,Teo:2010zb,Wang:2012bgb}
for the Maxwell-Amp\`ere law, 
which is closely related to  
the anomaly inflow mechanism of the axionic strings~\cite{Callan:1984sa,Naculich:1987ci}
(see also 
Refs.~\cite{AlvarezGaume:1983cs,Harvey:1988in,Kaplan:1987kh,Manohar:1988gv,Townsend:1993wy}).

On the other hand,
the notion of the symmetries has been recently
generalized to so-called higher $p$-form 
symmetries~\cite{Banks:2010zn,Kapustin:2014gua,Gaiotto:2014kfa}, which are
symmetries under transformations of 
$p$-dimensional extended objects 
(see also earlier Refs.~\cite{Batista:2004sc,Nussinov:2006iva,Nussinov:2008aa,Nussinov:2009zz,Nussinov:2011mz,Pantev:2005zs,Pantev:2005wj,Pantev:2005rh}
and related topics~\cite{Distler:2010zg}).
We can understand previously known phenomena
such as topologically ordered 
states with abelian anyons~\cite{Wen:1989iv,Wen:1990zza,Wen:1991rp,Hansson:2004wca} 
in terms of the higher-form symmetries.
Furthermore, 
these symmetries have revealed
new aspects of phases of matter for 
gauge theories~\cite{Gaiotto:2017yup,Gaiotto:2017tne,Tanizaki:2017qhf,Tanizaki:2017mtm,Komargodski:2017dmc,Hirono:2018fjr,Hirono:2019oup,Hidaka:2019jtv,Misumi:2019dwq,Anber:2019nze,Anber:2020gig,Furusawa:2020kro}.

Combining the above mentioned two topics together,
it may be a natural question 
whether there are any underlying 
higher-form symmetries 
in the axion electrodynamics, 
and if so, whether they offer 
fundamental tools to describe the Witten effect
and anomalous Hall effect.
In the case that either axion or photon can be  
regarded as a background field, 
higher-form symmetries have been studied in 
Refs.~\cite{Cordova:2018cvg,Cordova:2019uob,Cordova:2019jnf,Anber:2020xfk}.
When both the axion and photon 
are dynamical degrees of freedom,
the symmetry generators of the higher-form symmetries 
are deformed~\cite{Sogabe:2019gif}.
In such a case, however,
a relation between the higher-form symmetries
and the above mentioned effects 
is yet to be clarified.

In this paper, we study the higher-form symmetries of 
the massless 
axion electrodynamics in $(3+1)$ dimensions 
by using a low-energy effective theory 
described by
a massless axion and massless photon
with the topological coupling.
The Witten effect and anomalous Hall effect 
can be described by this effective theory, 
which can be model-independently 
determined only by symmetry.
We further show that the higher-form symmetries
are accompanied by a mathematical structure 
called as the semistrict 3-group 
or 2-crossed module 
in the mathematical literature~\cite{CONDUCHE1984155,Martins:2009evc}. 
Hereafter, we will simply refer to the semistrict 3-group as 
the 3-group.
We then find that the Witten effect and anomalous Hall effect
can be understood as symmetry transformations of the 3-group. 
Roughly speaking, the 3-group is a set of three groups with maps
between them,  
and is a generalization of an ordinary group 
and a 2-group (crossed module)~\cite{Baez:2002jn,Baez:2003fs,Baez:2004in,Baez:2005qu,Baez:2010ya,Ho:2012nt,Kapustin:2013uxa,Kapustin:2013qsa,Soncini:2014zra,Sharpe:2015mja,Bhardwaj:2016clt,deAlmeida:2017dhy,Cordova:2018cvg,Benini:2018reh,Delcamp:2018wlb,Wen:2018zux,Delcamp:2019fdp,Thorngren:2020aph}.
Since the explicit definition of the 3-group may be rather 
abstract compared with 
ordinary groups,
here we give a diagrammatic and intuitive explanation of the 3-group 
based on higher-form symmetries,
which are enough to understand underlying physics. 
In the forthcoming paper, we will show that 
this physical explanation satisfies all the axioms of 
the 3-group~\cite{HNY:inprep}.

The appearance of the 3-group in physics is not the first time;
It has been already applied to 
string theory, e.g.,
Aharony-Bergman-Jafferis-Maldacena model 
describing membranes
in M-theory~\cite{Aharony:2008ug,Martins:2009evc,Wang:2013dwa,Palmer:2013ena,Palmer:2014jma,Jurco:2016qwv,Samann:2019nuj} 
and to quantum gravity~\cite{Radenkovic:2019qme}.\footnote{Note that the 3-group in this paper
is different from a 3-Lie group based on a 
3-bracket~\cite{FAULKNER19731,Filippov:1986}.
The 3-Lie group has been applied to 
Bagger-Lambert-Gustavsson model
in M-theory~\cite{Bagger:2007jr,Gustavsson:2007vu,Palmer:2013pka,Palmer:2013ena,Palmer:2014jma}.
}
However, we emphasize that 
such a 3-group structure very naturally arises in a simpler and more familiar system, 
{\it i.e.}, the axion electrodynamics, 
and therefore there should be potential applications to both high energy and condensed matter physics.

This paper is organized as follows. 
In section \ref{s:setup}, we give an action of 
massless axion electrodynamics and gauge invariant observables.
The higher-form symmetries
of the system are shown in section \ref{s:hfsym}.
We discuss in section \ref{s:corr}
correlation functions of the symmetry generators.
We interpret the Witten effect 
and anomalous Hall effect around axionic strings 
in terms of the symmetry generators.
In section \ref{s:3g}, we discuss an underlying 3-group structure 
of the system with an intuitive introduction of the 3-group.
We relate the structure of the 3-group 
with the Witten effect and anomalous Hall effect.
Finally, we summarize this paper in section \ref{s:sum}.

\section{Setup\label{s:setup}}
We consider the following action~\cite{Wilczek:1987mv},
\begin{equation}
 S = -\int 
\(\tf{v^2}{2}  d\phi \wedge \star d\phi + \tf{1}{2e^2} da \wedge \star da 
- 
\tf{N}{8\pi^2} \phi  da \wed da \) .
\label{200601.0412}
\end{equation}
Here, $\phi$ is an axion, $a$ a photon, 
 $v$  a mass dimension 1 parameter, 
$e$  a coupling constant of the photon,
$N$ an integer, and $\star$ the Hodge star operator.
The axion $\phi$ is a pseudo-scalar field with 
a $2\pi$ periodicity at each point in the spacetime ${\cal P}$,
\begin{equation}
 \phi ({\cal P})+ 2\pi \sim \phi ({\cal P}). 
\label{200531.1745}
\end{equation}
We regard the periodicity as redundancy of the axion.
The redundancy can be understood as a gauge symmetry,
called a $(-1)$-form 
gauge symmetry~\cite{Cordova:2019uob,Cordova:2019jnf}.
By the identification, the axion can have a winding number
on a closed loop ${\cal C}$,
\begin{equation}
 \int_{\cal C} d\phi \in 2\pi \bb{Z}.
\label{200608.1327}
\end{equation}
The non-zero winding number implies that  a worldsheet of an axionic string
can be linked with ${\cal C}$.
An object invariant under the redundant transformation 
in \er{200531.1745} is a local object at a point ${\cal P}$,
\begin{equation}
 I(q_{\phi E}, {\cal P}) = e^{i q_{\phi E} \phi({\cal P})},
\label{200601.0648}
\end{equation}
where $q_{\phi E}$ should be an integer.

The photon $a$ is described by a $U(1)$
1-form gauge field, 
whose gauge transformation is given by
$a \to a+ d\lambda$ with a 0-form gauge parameter $\lambda$.
In addition, the $U(1)$ gauge field 
$a$ is subject to the Dirac quantization condition,
\begin{equation}
 \int_{\cal S} da \in 2\pi \bb{Z},
\label{200608.1328}
\end{equation}
where ${\cal S}$ is a closed surface 
(e.g., a sphere $S^2$).
The non-zero value of $ \int_{\cal S} da$ implies 
that there can be a magnetic monopole in the interior of 
${\cal S}$.
A gauge invariant object given by $a$ is a Wilson loop, 
\begin{equation}
 W(q_{aE}, {\cal C}) = e^{iq_{aE} \int_{\cal C} a},
\label{200601.1602}
\end{equation}
where $q_{aE}$ should be an integer as a result of the Dirac quantization condition.

\section{Higher-form global symmetries\label{s:hfsym}}

In this system, there are higher-form global symmetries
in addition to ordinary ones.
In general, a $p$-form symmetry in $D$ spacetime dimensions 
is a symmetry whose 
charged object is a $p$-dimensional object.
The corresponding symmetry generator is given by 
a topological object labeled by a group element on a $(D-p-1)$-dimensional subspace,
which satisfies the group composition rule.
The term `topological' means that the subspace of
 the object can be continuously deformed as long as 
it does not intersect with the charged object.
The symmetry transformation of the charged object is 
generated by the linking of the charged object and the symmetry generator.

We now establish the higher-form symmetries of the 
system in \er{200601.0412}.
We show the symmetry generators related to
equations of motion (EOM) and Bianchi identities,
and we specify symmetry groups compatible with the quantization conditions 
in \ers{200608.1327} and \eqref{200608.1328}.

\subsection{$\bb{Z}_N$ 0-form symmetry}
$\bb{Z}_N$ 0-form symmetry is 
a shift symmetry of the axion. 
It is
obtained by 
the EOM of the axion,
$ v^2 d\star d\phi + \tf{N}{8\pi^2} da \wed da =0$,
which leads to the following 
closed current 3-form,
\begin{equation}
\begin{split}
 j_{\phi E} 
& = 
- v^2 \star d\phi - \tf{N}{8\pi^2} a \wed da.
\end{split}
\end{equation}
The corresponding charge is given by an integral
on a closed 3-dimensional subspace 
${\cal V}$,
\begin{equation}
 Q_{\phi E} ({\cal V})= \int_{\cal V} j_{\phi E}.
\end{equation}
Note that we can obtain an ordinary conserved charge 
by choosing ${\cal V} $ as a time slice $\bb{R}^3$.
The charge is topological meaning that 
$Q_{\phi E} ({\cal V})$ does not change 
under a small deformation of 
${\cal V}$ by a 4-dimensional subspace $\Omega$, 
${\cal V} \to {\cal V}+ \der \Omega$:
$ Q_{\phi E} ({\cal V}+\der \Omega)-Q_{\phi E} ({\cal V})
= 
 \int_{\der \Omega} j_{\phi E}
=
\int_{\Omega} dj_{\phi E}
=0,
$
where we have used the Stokes theorem.

The symmetry generator made of $j_{\phi E}$ is 
given by 
\begin{equation}
 U_{\phi E}(e^{2\pi i n_\phi/N},{\cal V})
 = e^{\tf{2\pi i n_\phi}{N} Q_{\phi E}({\cal V})}.
\end{equation}
Here, 
$e^{2\pi i n_\phi /N} \in \bb{Z}_N$ 
parameterizes the generator.
One might think that the parameter is a $U(1)$-valued 
continuous number, $e^{i\alpha_{\phi E}} \in U(1)$, 
since a conserved current would be related to 
a continuous symmetry.
However, the Dirac quantization condition requires that 
the parameter is a $\bb{Z}_N$-valued number.
This is due to the gauge variant term 
$\tf{N}{8\pi^2} a \wed da $ in $j_{\phi E}$.
In order to make the term 
$\exp( i\tf{N\alpha_{\phi E}}{8\pi^2} \int_{\cal V} a \wed da)$ 
be consistent with the Dirac quantization, 
$e^{i\alpha_{\phi E}}$ should satisfy $e^{i\alpha_{\phi E} } \in \bb{Z}_N$.
This requirement is the same as the quantization of 
the coefficient of a Chern-Simons term in $(2+1)$ 
dimensions~\cite{Henneaux:1986tt}.

The topological object $ U_{\phi E}(e^{2\pi i n_\phi /N},{\cal V})$ 
generates the $\bb{Z}_N$ transformation on the local object 
$I(q_{\phi E}, {\cal P})$ in \er{200601.0648}.
In the path integral formulation, it can be expressed as a
correlation function,
\begin{equation}
\begin{split}
& \vevs{U_{\phi E}(e^{2\pi i n_\phi/N},{\cal V})
I(q_{\phi E}, {\cal P})} 
\\
& 
= e^{2\pi i n_\phi q_{\phi E}  \link ({\cal V},{\cal P})/N}
\vevs{I(q_{\phi E}, {\cal P})}.
\end{split}
\end{equation}
Here, the symbol `$\vevs{}$' denotes the vacuum expectation value (VEV), 
and 
$\link({\cal V,P}) \in \bb{Z}$ 
denotes the linking number between 
${\cal V} $ and ${\cal P}$.
The correlation function can be derived by the 
Schwinger-Dyson equation 
$\vevs{\delta_\phi{I(q_{\phi E}, {\cal P})} } 
=- i \vevs{(\delta_\phi{S})I(q_{\phi E}, {\cal P})}$.
Here $\delta_\phi$ denotes the variation with respect to $\phi$. 
Note that the correlation function is obtained as a 
normal ordered product that subtracts trivial divergences.
The 3-dimensional symmetry generator
$U_{\phi E}(e^{2\pi i n_{\phi}/N},{\cal V})$ 
acts on the 0-dimensional object $I(q_{\phi E}, {\cal P})$
as a $\bb{Z}_N$ transformation.
Therefore, it is identified as a $\bb{Z}_N$ 0-form symmetry.

We can regard $U_{\phi E}(e^{2\pi i n_{\phi }/N},{\cal V})$ 
as an external domain wall
if we take ${\cal V}$ as a spatially and temporally extended object.
This is because $U_{\phi E}(e^{2\pi i n_{\phi }/N},{\cal V})$
changes the value of $\phi$ to $\phi + \tf{2\pi n_{\phi}}{N}$.
\subsection{Electric $\bb{Z}_N$ 1-form symmetry}

Next, we show that there is a $\bb{Z}_N$ 1-form symmetry 
related to the EOM of $a$. 
We will call it an electric $\bb{Z}_N$ 1-form symmetry,
since the symmetry is related to the conservation of the electric flux.
The EOM of the photon,
$ \tf{1}{e^2} d\star da - \tf{N}{4\pi^2} d\phi \wed da =0,$
 leads to the conservation of the electric flux modified by the axion.
The corresponding closed current, charge and symmetry generator are
\begin{equation}
\begin{split}
 &j_{aE}
=
  \tf{1}{e^2} \star da -\tf{N}{4\pi^2} \phi da,\quad
 Q_{aE} ({\cal S})
 = \int_{\cal S} j_{aE},
\\
& U_{aE} (e^{2\pi i n_a/N}, {\cal S})
 = e^{\tf{2\pi i n_a}{N}Q_{aE} ({\cal S})},
\end{split}
\end{equation}
respectively.
The charge $Q_{aE}({\cal S})$ is topological under a 
small deformation of ${\cal S}$: 
${\cal S} \to {\cal S} + \der {\cal V}_0$,
where ${\cal V}_0 $ is a 3-dimensional subspace
with a boundary.
The parameter $e^{2\pi i n_a/N}$ should again 
be a $\bb{Z}_N$ valued 
number in order to make $U_{aE}$ be invariant under 
the redundant transformation of $\phi$ in \er{200531.1745}.

The Wilson loop in \er{200601.1602} is a 
gauge invariant charged object 
transformed by $U_{aE}$. 
The Schwinger-Dyson equation 
$\vevs{\delta_a W(q_{aE}, {\cal C})}
= -i\vevs{(\delta_a S) W(q_{aE}, {\cal C})}$ 
leads to 
\begin{equation}
\begin{split}
& \vevs{U_{aE}(e^{2\pi i n_a/N},{\cal S})
W(q_{aE}, {\cal C})}
\\
& = e^{2\pi i n_a q_{aE} \link({\cal S, C})/N}
\vevs{W(q_{aE}, {\cal C})},
\end{split}
\end{equation}
where $\delta_a$ denotes the variation with respect to $a$.
Since $e^{2\pi i n_a/N} \in \bb{Z}_N$,
the correlation function can be understood as a $\bb{Z}_N$
transformation.
Since the Wilson loop is a 1-dimensional object, 
the corresponding symmetry is a $\bb{Z}_N$ 1-form symmetry.

We can regard $U_{a E}(e^{2\pi i n_{a}/N},{\cal S})$ 
as an external electric flux localized on and 
perpendicular to ${\cal S}$
if we take ${\cal S}$ as an instantaneous surface.
We can also regard it 
as an external magnetic flux parallel to a time slice of ${\cal S}$
if we take ${\cal S}$ as a spatially and temporally 
extended surface.
\subsection{Magnetic $U(1)$ 1-form symmetry}

The gauge field $a$ obeys the Bianchi identity $d d a =0$,
which implies the conservation of the magnetic flux.
The identity gives us the following 2-form closed current,
 conserved charge, and symmetry generator,
\begin{equation}
\begin{split}
& j_{a M} = \tf{1}{2\pi} da, 
\quad
Q_{aM} ({\cal S}) = \int_{\cal S} j_{aM},
\\
&U_{aM}(e^{i\alpha_{a}}, {\cal S}) = e^{i\alpha_{a} Q_{aM}({\cal S})},
\end{split}
\end{equation}
respectively.
Here, $e^{i\alpha_{a}} $ is a $U(1)$ valued number.
The charge $Q_{aM}$ is again a topological object.
The normalization of the current is determined by 
the Dirac quantization.

The charged object is a 't Hooft loop $T(q_{aM}, {\cal C}_{\rm mon.})$
that is a 
closed worldline of a magnetic monopole given by a charge $q_{aM}\in \bb{Z}$
and a loop ${\cal C}_{\rm mon.}$.
In the presence of the magnetic monopole, the Gauss law
of the magnetic flux is changed as
$Q_{aM}({\cal S}) = q_{aM} \link ({\cal S,C}_{\rm mon.})$.
In terms of the correlation function, $U_{aM}$ has a 
non-trivial phase in the presence of the 't Hooft loop:
\begin{equation}
\begin{split}
& \vevs{U_{aM}(e^{i\alpha_{a}}, {\cal S}) T(q_{aM}, {\cal C}_{\rm mon.})}
\\
& 
= e^{i\alpha_{a} q_{aM} \link ({\cal S,C}_{\rm mon.}) }\vevs{T(q_{aM}, {\cal C}_{\rm mon.})}. 
\end{split}
\end{equation}

By this correlation function,
$U_{aM}$ generates a $U(1)$ transformation on the 't Hooft loop.
The corresponding symmetry is a $U(1)$ 1-form symmetry,
since the 't Hooft loop is a 1-dimensional object.
We call this symmetry a magnetic $U(1)$ 1-form symmetry,
since it is related to the conservation of the magnetic flux.

\subsection{$U(1)$ 2-form symmetry}
Finally, we introduce a $U(1)$ 2-form symmetry.
The symmetry is originated from the Bianchi identity of the 
axion, $dd\phi =0$.
The closed current 1-form, conserved charge, and symmetry generator
are given by
\begin{equation}
\begin{split}
& j_{\phi M} = \tf{1}{2\pi} d\phi, 
\quad
Q_{\phi M} ({\cal C}) = \int_{\cal C} j_{\phi M}, 
\\
&
U_{\phi M} (e^{i\alpha_{\phi}}, {\cal C})
= e^{i \alpha_{\phi }Q_{\phi M} ({\cal C})}, 
\end{split}
 \end{equation}
respectively.
The charged object is a closed worldsheet of an axionic string
$V(q_{\phi M}, {\cal S}_{\rm str.})$ with a winding number $q_{\phi M}$.
In the presence of the axionic string, $Q_{\phi M} ({\cal C})$
has non-zero value, 
$Q_{\phi M} ({\cal C}) = q_{\phi M} \link({\cal C,S}_{\rm str.})$.
In terms of the correlation function of the symmetry generator, 
it can be written as
\begin{equation}
\begin{split}
& \vevs{U_{\phi M}(e^{i\alpha_{\phi}}, {\cal C})
V(q_{\phi M},{\cal S}_{\rm str.})}
\\
&
= e^{i\alpha_{\phi}q_{\phi M} \link({\cal C,S}_{\rm str.})}
\vevs{V(q_{\phi M},{\cal S}_{\rm str.})}. 
\end{split}
\end{equation}
The correlation function can be understood as a $U(1)$ 
symmetry transformation 
on a 2-dimensional object.
Therefore, the symmetry is a $U(1)$ 2-form symmetry.

\section{Correlations of symmetry generators 
\label{s:corr}}

In this section, we show that the symmetry generators 
are
correlated to themselves.
\begin{table*}[t]
{\small 
\begin{tabular}[t]{c||c|c|c|c}
\hline
Correlation function 
& 
$U_{\phi E}(e^{2\pi i n'_\phi/N }, {\cal V}_2)$ 
& $U_{aE}(e^{2\pi i n_a'/N}, {\cal S}_{E 2})$ 
& $U_{aM} (e^{i\alpha'_{a}},{\cal S}_{M 2})$
 & $U_{\phi M}(e^{i\alpha'_\phi}, {\cal C}_2)$
\\
\hline
\hline
$U_{\phi E} (e^{2\pi i n_\phi /N}, {\cal V}_1)$ 
& 
1 
& 
$U_{a M}(e^{-2\pi i n_\phi n_a' /N}, {\cal S}_{E 2} 
\cap \Omega_{{\cal V}_1})$
 &
1
  & 
1
\\
\hline
$U_{aE}(e^{2\pi i n_a/N}, {\cal S}_{E 1})$ 
&
$U_{aM}(e^{-2\pi i n_a n'_\phi /N}, {\cal V}_2 \cap {\cal V}_{{\cal S}_{E1}})$
&
$U_{\phi M}(e^{-2\pi i n_a n'_a /N}, {\cal V}_{{\cal S}_{E1}} \cap {\cal S}_{E2})$
&
1
&
1
\\
\hline
$U_{aM} (e^{i\alpha_{a}},{\cal S}_{M1})$
 &
1&1&1&1
\\
\hline
$U_{\phi M}(e^{i\alpha_\phi}, {\cal C}_1)$ 
&1&1&1&1
\\
\hline
\end{tabular}
}
\caption{
The correlation functions of the symmetry generators.
${\cal V}_{1,2}$, ${\cal S}_{E1,2}$, ${\cal S}_{M1,2}$, 
and ${\cal C}_{1,2}$ are 
3-, 2-, 2-, and 1-dimensional closed subspaces.
The object
$A = U_{\phi E}(e^{2\pi i n_\phi/N }, {\cal V}_1),
..., U_{\phi M}(e^{i\alpha_\phi}, {\cal C}_1)$ 
in the left columns  
acts on the one 
$B = U_{\phi E}(e^{2\pi i n'_\phi/N }, {\cal V}_2),
..., U_{\phi M}(e^{i\alpha'_\phi}, {\cal C}_2)$
 in the top row.
The object $C$ in the correlation function 
$\vevs{AB} = \vevs{CB}$ is displayed in 
the rest of the table.\label{tab:corr}} 
\end{table*}
The correlation can be basically evaluated by 
the Schwinger-Dyson equations.
By the correlation functions, we can understand some phenomena 
of the axion such as the Witten effect 
and 
the anomalous Hall effect in terms of symmetry transformations.
We will use the correlation functions of the 
symmetry generators to show a 3-group structure in this system.

\subsection{Action of 0-form symmetry generator 
on electric 1-form symmetry generator
as Witten effect}
First, we show that the generator of the 0-form symmetry 
can act on the generator of the electric 1-form symmetry, and 
gives the generator of the magnetic 1-form symmetry.
The Schwinger-Dyson equation,
 $\vevs{\delta_\phi j_{aE} }
= -i\vevs{(\delta_\phi{S})j_{aE}}$,
would give us
$ i \vevs{Q_{\phi E} ({\cal V}) Q_{aE}({\cal S}) }
=
-\tf{N}{2\pi} \vevs{Q_{aM} ({\cal S}\cap \Omega_{\cal V})}$,
where $\Omega_{\cal V}$ is a 4-dimensional subspace 
satisfying $\der \Omega_{\cal V} = {\cal V}$.
However, the generator of the 
left-hand side is meaningful mod $N$.
The relation that is invariant under the redundancy is 
the correlation function of the symmetry generators:
\begin{equation}
\begin{split}
& \vevs{
U_{\phi E}
(e^{2 \pi i n_\phi /N},{\cal V})
 U_{aE}(e^{2\pi i n_{a}/N},{\cal S})}
\\
&=
\vevs{
U_{aM}(e^{ - 2 \pi i n_{a}n_{\phi}/N} ,
{\cal S}\cap \Omega_{\cal V})
 U_{aE}(e^{2\pi i n_{a}/N},{\cal S}) }. 
\end{split}
\label{200605.0125}
\end{equation}

\begin{figure}[t]
\includegraphics[width=25em]{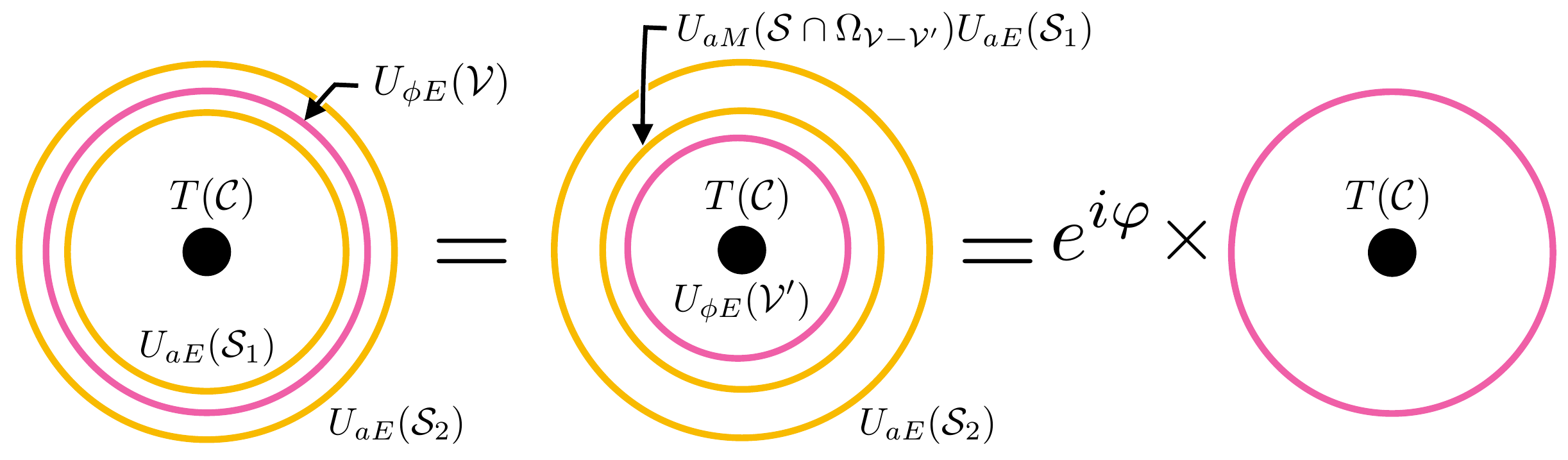}
\caption{
\label{f:witten-sikivie}  
The correlation function in \er{200602.2124}.
This is the figure of a time slice of the correlation function.
The black dot represents a monopole, ${\cal V}$ a time of domain wall, 
${\cal S}={\cal S}_1-{\cal S}_2$ an instantaneous surface,
and $\Omega_{{\cal V}-{\cal V'}}$ the space between ${\cal V}$ and ${\cal V'}$.
The left hand side counts the difference of the electric flux between
$T(q_{aM}, {\cal C})$ and $U_{\phi E}(e^{2\pi i n_\phi},{\cal V}_2) T(q_{aM}, {\cal C})$
by $U_{aE}(e^{2\pi i n_a},{\cal S})$. 
The middle side counts the magnetic flux from 
$T(q_{aM},{\cal C})$ by 
$U_{a M}(e^{2\pi i n_\phi n_a},{\cal S}\cap \Omega_{{\cal V}-{\cal V}'})$,
where we used ${\cal S}_1\cap \Omega_{{\cal V}-{\cal V}'}={\cal S}\cap \Omega_{{\cal V}-{\cal V}'}$.
The right panel shows the evaluation of the diagram in the middle panel.
The obtained phase is $\varphi = -2\pi  n_{a}n_{\phi}q_{aM}
\link({\cal S}\cap \Omega_{{\cal V}-{\cal V'}}, {\cal C})/N$.} 
\end{figure}

We now argue that this is nothing but the Witten effect~\cite{Witten:1979ey} due to 
the presence of the axion~\cite{Fischler:1983sc}.
The Witten effect is an effect on a magnetic monopole
in the presence of a $\theta$-term in electromagnetism
$\tf{\theta}{8\pi^2} da \wed da$.
This term induces 
an electric charge $\tf{\theta}{2\pi}$ for the monopole.
The effect can also arise in the presence of non-zero value of the axion.
In our case, the Witten effect can be expressed by 
the insertion of the 't Hooft loop:
\begin{equation}
\begin{split}
& \vevs{
 U_{aE}(e^{2\pi i n_a/N},{\cal S})
 U_{\phi E}
(e^{2\pi i n_\phi /N },{\cal V})
 T(q_{aM},{\cal C})}
\\
&=
\vevs{
  U_{aE}(e^{2\pi i n_a/N},{\cal S})U_{aM}(e^{-2\pi i n_{a}n_{\phi}/N } ,
{\cal S}\cap \Omega_{{\cal V}-{\cal V'}})\\
&\qquad\times U_{\phi E}(e^{2\pi i n_\phi /N },{\cal V'}) T(q_{aM},{\cal C})}
\\
&=
e^{i\varphi}\vevs{ U_{\phi E}(e^{2\pi i n_\phi /N },{\cal V})  T(q_{aM},{\cal C})},
\end{split} 
\label{200602.2124}
\end{equation}
where $\varphi = -2\pi  n_{a}n_{\phi}q_{aM} 
\link({\cal S}\cap \Omega_{{\cal V}-{\cal V'}}, {\cal C})/N$.
We measure the difference of the electric flux between 
$T$ and $U_{\phi E} T$ by $U_{aE}(e^{2\pi i n_a/N},{\cal S})$ (see Fig.~\ref{f:witten-sikivie} for the configuration).
We take $U_{aE}(e^{2\pi i n_a/N},{\cal S})\to 1$ and $U_{\phi E}(e^{2\pi i n_\phi /N },{\cal V'})\to U_{\phi E}(e^{2\pi i n_\phi /N },{\cal V})$ in the last line, thanks to the topological nature of symmetry generators.
The correlation means that the domain wall wrapping the monopole
is a source of the electric flux.
There is non-zero surface electric charge density 
if a magnetic flux transversally 
passes through an axionic domain wall,
and the electric flux is determined by the original magnetic flux~\cite{Sikivie:1984yz}. 
This domain wall enclosing the monopole is 
called a ``monopole bag''~\cite{Kogan:1992bq,Kogan:1993yw}.

\subsection{Correlation of two 1-form symmetry generators
 as anomalous Hall effect
}
Second, we evaluate the correlation function of 
the 1-form symmetry generators, and give 
the 2-form symmetry generator.
We then show that it implies 
the anomalous Hall effect around
an axionic string.
The Schwinger-Dyson equation 
$\vevs{\delta_aj_{aE}}
= -i\vevs{(\delta_a{S}) j_{aE}}$
gives us 
\begin{equation}
\begin{split}
&\vevs{
U_{aE} (e^{2\pi i n_a/N},{\cal S}_1)
U_{aE} (e^{2\pi i n'_a/N},{\cal S}_2)} 
\\
&
= \vevs{U_{\phi M} (e^{-2\pi i n_an_a'/N},{\cal S}_2 
\cap {\cal V}_{{\cal S}_1})
U_{aE} (e^{2\pi i n'_a/N},{\cal S}_2)}, 
\end{split}
\label{200603.0232}
\end{equation}
where ${\cal V}_{{\cal S}_1}$ 
is a 3-dimensional subspace whose boundary 
is ${\cal S}_1$, $\der {\cal V}_{{\cal S}_1} ={\cal S}_1$.
In the above correlation function, 
we have regularized a trivial divergence. 
Note that the correlation function of the charges would be 
$ i\vevs{Q_{aE}({\cal S}_1)Q_{aE} ({\cal S}_2)}
 = -\tf{N}{2\pi} \vevs{Q_{\phi M} ({\cal S}_2\cap {\cal V}_{{\cal S}_1})}
$,
but the left hand side is meaningful mod $N$. 

\begin{figure}[t]
 \begin{center}
  \includegraphics[width=25em]{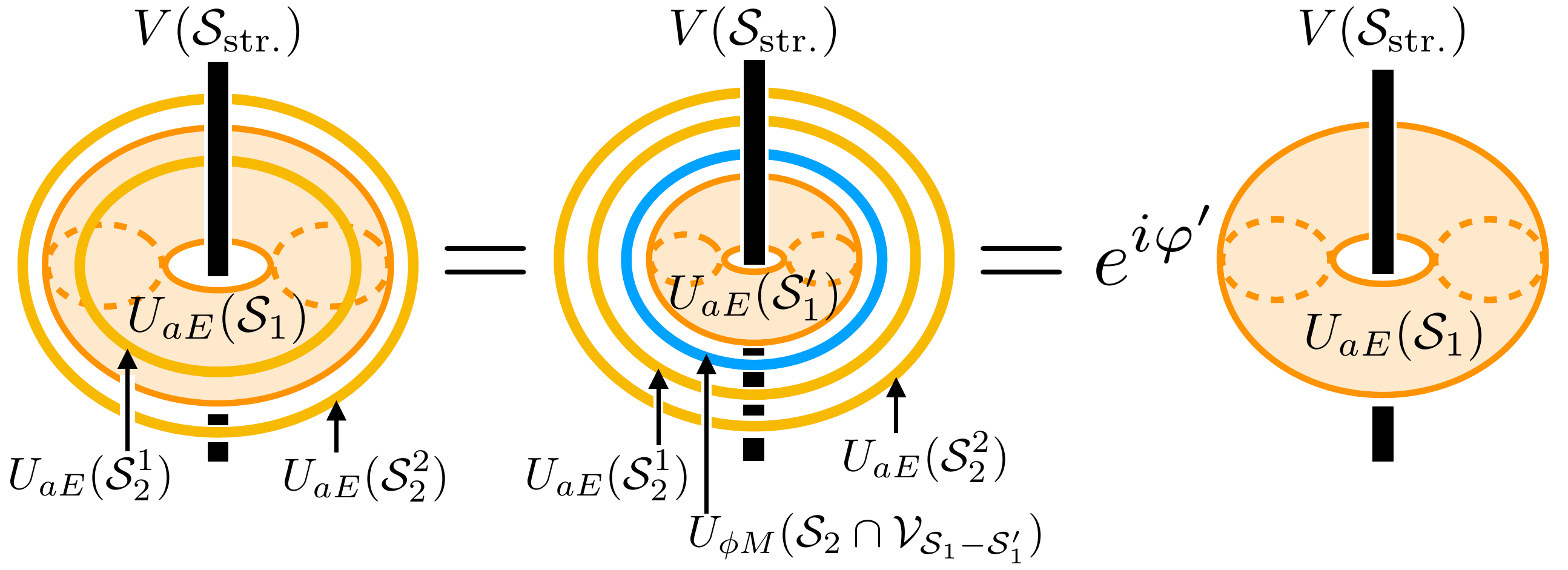}
 \end{center}
\caption{\label{f:AHE}The configuration of in \er{200603.0238}.
This is the figure of a time slice of the correlation function.
The black line represents an axionic string, 
${\cal S}_1$ an instantaneous torus surface of electric flux, 
${\cal S}_2={\cal S}^1_2-{\cal S}^2_2$ a time of 
surface operator that detects the magnetic flux,
${\cal V}_{{\cal S}_1-{\cal S}_1'}$ 
the space between ${\cal S}_1$ and ${\cal S}'_1$.
The time slice of 
${\cal S}^1_2$ is in the interior of the torus ${\cal S}_1$.
The left hand side counts the difference of the magnetic flux 
between
$V(q_{\phi M}, {\cal S}_{\rm str.})$ 
and $U_{a E}(e^{2\pi i n_a},{\cal S}_1) V(q_{\phi M}, {\cal S}_{\rm str.})$
by $U_{aE}(e^{2\pi i n'_a},{\cal S}_2)$. 
The middle side counts the winding number due to 
the axionic string  
$V(q_{\phi M}, {\cal S}_{\rm str.})$ by 
$U_{\phi M}(e^{-2\pi i n_a n'_a},{\cal S}_2 
\cap {\cal V}_{{\cal S}_1-{\cal S}_1'})$,
where we used 
${\cal S}^1_2\cap 
{\cal V}_{{\cal S}_1-{\cal S}_1'}
={\cal S}_2\cap {\cal V}_{{\cal S}_1-{\cal S}_1'}$.
The right panel shows the evaluation of the diagram in the middle panel.
The obtained phase is 
$\varphi'= -2\pi n_an_a'q_{\phi M} \link({\cal S}_2 \cap {\cal V}_{{\cal S}_1-{\cal S}_1'}, {\cal S}_{\rm str.})/N$.
}
\end{figure}

We argue that the correlation function \er{200603.0232} 
means the 
anomalous Hall effect around the axionic string.
There is a non-zero electric current 
in the presence of the axionic string and electric flux.
The direction of the current is 
perpendicular to both of the electric flux 
and the gradient of the axion.
In our case, 
the anomalous Hall current arises 
on the surface ${\cal S}_1$, 
and we can measure the magnetic flux 
due to the current
 by the Amp\`ere law.
In terms of the correlation function,
it can be written 
by the insertion of the
worldsheet of the axionic string $V(q_{\phi M}, {\cal S}_{\rm str.})$:
\begin{equation}
\begin{split}
&
\vevs{
U_{aE} (e^{2\pi i n'_a/N},{\cal S}_2)
U_{aE} (e^{2\pi i n_a/N},{\cal S}_1)
V(q_{\phi M}, {\cal S}_{\rm str.})} 
\\
&
= 
\vevs{
U_{aE} (e^{2\pi i n'_a/N},{\cal S}_2)
U_{\phi M} (e^{-2\pi i n_an_a'/N},{{\cal S}_2 \cap 
{\cal V}_{{\cal S}_1-{\cal S}_1'}})
\\
&\qquad
\hph{\langle}
\times 
U_{aE} (e^{2\pi i n_a/N},{\cal S}'_1)
V(q_{\phi M}, {\cal S}_{\rm str.})} 
\\
&
= 
e^{i\varphi'}
\vevs{
U_{aE} (e^{2\pi i n_a/N},{\cal S}_1)
V(q_{\phi M}, {\cal S}_{\rm str.})},
\end{split}
\label{200603.0238}
\end{equation}
where 
$\varphi'= -2\pi n_an_a'q_{\phi M} \link({\cal S}_2 \cap {\cal V}_{{\cal S}_1-{\cal S}_1'}, {\cal S}_{\rm str.})/N.$
The configuration is shown in Fig.~\ref{f:AHE}.
In the last line, we have applied topological deformations
$U_{aE} (e^{2\pi i n'_a/N},{\cal S}_2) \to 1$ 
and $U_{aE} (e^{2\pi i n_a/N},{\cal S}'_1) \to U_{aE} (e^{2\pi i n_a/N},{\cal S}_1)$.
Physically, the correlation means that 
we measure the difference of the magnetic flux 
between the interior and exterior of ${\cal S}_1$.

Using the Schwinger-Dyson equation, one can show that other correlation functions between symmetry generators are trivial.
We summarize the correlation functions in Table~\ref{tab:corr}.

\section{3-group structure in axion electrodynamics\label{s:3g}}

By the above discussions, we here show that 
the symmetry generators of the system obey a 3-group.
The 3-group is a set of three groups denoted as $G$, $H$, and $L$,
an action of $G$ on the groups, and the so-called Peiffer lifting.
The mathematical definition of the 3-group is 
given in Ref.~\cite{CONDUCHE1984155} (See also the 
literature in physics~\cite{Palmer:2013pka,Palmer:2013ena,Palmer:2014jma,Radenkovic:2019qme}.)

Instead, we here give an intuitive introduction of the 3-group
in terms of the higher-form symmetries.
We need only two of the axioms of the 3-group
that are sufficient for our discussion.
The other axioms are automatically satisfied
in our case, since the symmetry generators do not have boundaries.
We will argue that the following intuitive definitions 
satisfy all of the axioms of the 3-group
in a forthcoming paper~\cite{HNY:inprep}.

One of the axioms is the action of $G$.
$G$ can act on $G$, $H$, and $L$ by automorphisms,
which is a generalization of conjugation.
The action of $g \in G$ on the elements 
$g' \in G$, $h \in H$, and 
$l \in L$ are denoted as 
$g \trr g'$, $g \trr h$, and $g \trr l$.
The action $g \trr g'$ is defined by conjugation,
$g \trr g':= gg'g^{-1}$.

In terms of the higher-form symmetries, 
the groups $G$, $H$, and $L$ are symmetry groups of 
0-, 1-, and 2-form symmetries.
The 0-form symmetry generator can act on the 0-, 1-, and 2-form 
symmetry generators by enclosing them.

Another axiom is that 
there is a map from two elements of $H$ to $L$,
denoted as $\{h,h'\} \in L$ for $h,h'\in H$.
The curly bracket $\{-,-\}$ is called the Peiffer lifting.
The Peiffer lifting should be compatible with the action of $G$, 
$g\trr (\{h, h'\}) = \{g \trr h, g \trr h'\}$.

In the viewpoint of the higher-form symmetries, 
the Peiffer lifting corresponds to 
the linking (braiding) of the two 1-form symmetry generators 
$H$ leading to a 2-form symmetry generator $L$.
The compatibility with the action corresponds to 
the fact that we can freely insert a trivial 0-form symmetry generator 
$1 = g^{-1}g$ in the linking of two 1-form symmetry generators.

Let us identify the action $\trr$ and the Peiffer lifting $\{-,-\}$
in the axion-photon system.
In our system, the groups are identified as 
$G = \bb{Z}_N$ for the 0-form symmetry, 
$H = \bb{Z}_N \times U(1)$ for the electric 
and magnetic 1-form symmetries, 
and $L = U(1)$ for the 2-form symmetry.

\begin{figure}[t]
\begin{center}
\includegraphics[height=6em]{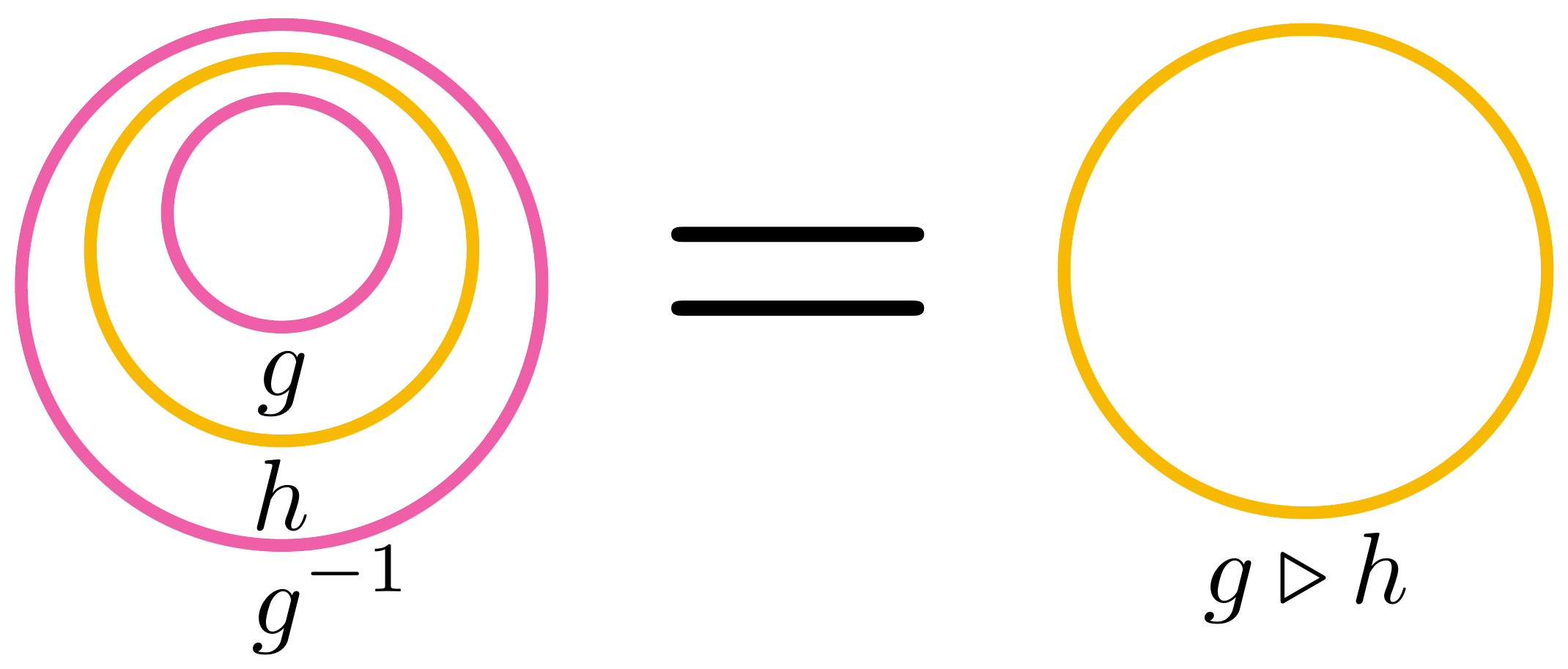} 
\end{center}
\caption{\label{f:action} The action of $G$ on $H$.
This is a figure of a time and space slice of the objects.
$g$ and $g^{-1}$ are 3-dimensional objects extended to 
temporal and spatial directions, e.g., $\bb{R}\times S^2$.
$h$ is an instantaneous surface object, e.g., $S^2$,
and is put between $g$ and $g^{-1}$.}
\end{figure}

\begin{figure}[t]
\begin{center}
 \includegraphics[height=6em]{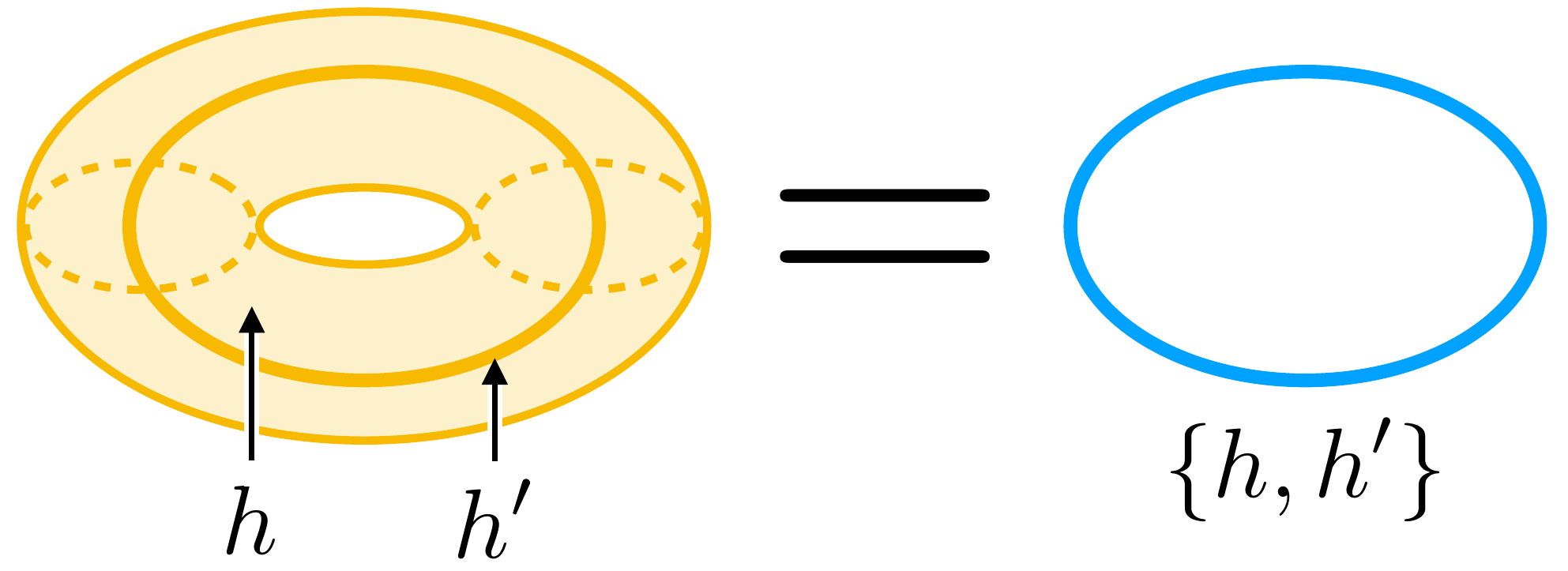} 
\end{center}
\caption{\label{f:peiffer} Peiffer lifting:
This is a figure of a time slice of the objects.
Both of $h$ and $h'$ are surface objects.
$h$ is an instantaneous surface of a torus. 
$h'$ is, e.g., 
a torus extended to
temporal and spatial directions,
and one circle of the 
time slice of the torus of $h'$ is in the interior of 
the torus of $h$.
The other circle of the time slice of $h'$ is in the exterior of $h$,
and is omitted. 
$\{h,h'\}$ is a line object, 
which is an instantaneous circle.}
\end{figure}

Let us first see the action of $G$ on $G$, $H$, and $L$.
From the first line in Table~\ref{tab:corr}, 
the action on $G$ and $L$ is trivial: 
$e^{2\pi i n_\phi/N} \trr e^{2\pi i n_\phi'/N} = e^{2\pi i n_\phi'/N}$, 
and $e^{2\pi i n_\phi} \trr e^{i\alpha_\phi} = e^{i\alpha_\phi}$.
On the other hand, the action on $H$ (Fig.~\ref{f:action})
is nontrivial:
$(e^{2\pi i n_a/N}, e^{i\alpha_a}) \in \bb{Z}_N \times U(1) $
is defined by
\begin{equation}
\begin{split}
& e^{2\pi i n_\phi/N} \trr (e^{2\pi i n_a/N}, e^{i\alpha_a}) 
=(e^{2\pi i n_a/N}, e^{-2\pi i n_a n_\phi /N} e^{i\alpha_a}) . 
\end{split}
\end{equation}
Second, we identify the Peiffer lifting.
The corresponding correlation function is 
given in \er{200603.0232},
because the correlation of the 
two of the 1-form symmetry generators 
leads to the 2-form symmetry generator.
The group parameter of the 2-form symmetry generator 
$U_{\phi M} (e^{-2\pi i n_an_a'/N},{\cal S' \cap V_S})$
gives us the definition of 
the Peiffer lifting as 
\begin{equation}
\{(e^{2\pi i n_a/N}, e^{i\alpha_a}), 
(e^{2\pi i n'_a/N}, e^{i\alpha'_a})\}
= e^{-2\pi i n_an_a'/N} \in  L.
\end{equation}
The Peiffer lifting is compatible with 
the action of $e^{2\pi i n_\phi/N} \in G$, 
because 
\begin{eqnarray}
&&\hspace{-1cm}
\{e^{2\pi i n_\phi/N} \trr (e^{2\pi i n_a/N}, e^{i\alpha_a}),
e^{2\pi i n_\phi/N}  \trr (e^{2\pi i n'_a/N}, e^{i\alpha'_a})\}
\nonumber \\
&&=
e^{-2\pi i n_an_a'/N},
\end{eqnarray}
and 
$e^{2\pi i n_\phi/N} \trr e^{-2\pi i n_an_a'/N} =  e^{-2\pi i n_an_a'/N} $.
Therefore, the higher-form symmetry groups of the 
axion-photon system
satisfy the axioms of the 3-group
by the above definitions of the action $\trr$ and 
the Peiffer lifting $\{-,-\}$.
The Peiffer lifting can be schematically shown in Fig.~\ref{f:peiffer}.
The linking of two tori in Fig.~\ref{f:peiffer} is called 
a surface link or 
Hopf 2-link in the mathematical literature~\cite{Carter:2001,NAKAMURA:2014}.

By the above discussion, we can now understand 
the Witten effect and anomalous Hall effect in the viewpoint of symmetries.
These phenomena are consequences of the 3-group structure:
the former is the action of $G$ on $H$, 
and the latter is the Peiffer lifting. 

One can discuss the spontaneous symmetry breaking (SSB) of the 
3-group. 
The SSB of the higher-form symmetry can be judged by 
the infra-red behavior of the VEV of the charged objects.
Since the axion is massless, the 
non-zero VEV of $\vevs{e^{i\phi ({\cal P})} e^{-i\phi ({\cal P'})}}$
for $|{\cal P}-{\cal P'}|\to \infty$
means that 
 $\bb{Z}_N$ 0-form symmetry is spontaneously broken.
Similarly, the non-zero VEVs of the 
Wilson loop and the 't Hooft loop lead to the SSB 
of the $\bb{Z}_N$ electric and $U(1)$ magnetic 1-form symmetries, 
respectively.
Finally, the massless axion prevents the SSB of the 
$U(1)$ 2-form symmetry, which is a higher-form generalization 
of the Coleman-Mermin-Wagner-Hohenberg theorem~\cite{Coleman:1973ci,Mermin:1966fe,Hohenberg:1967zz,Gaiotto:2014kfa,Lake:2018dqm}.

\section{Summary and discussion}
\label{s:sum}
In this paper, we have shown that the higher-form symmetries 
of an axion-photon system in $(3+1)$ dimensions
have a 3-group structure.
We have considered a massless axion with a topological coupling 
with a photon characterized by an integer $N$.
In this system, it has been shown that there are 
a $\bb{Z}_N$ 0-form symmetry, $\bb{Z}_N$ electric 1-form symmetry,
$U(1)$ magnetic 1-form symmetry, and $U(1)$ 2-form symmetry. 
We have evaluated the correlation functions of the 
symmetry generators.
It has been found that the correlation function of
0-form and electric 1-form symmetry generators leads to  
the magnetic 1-form symmetry generator.
We have understood this result as the Witten effect
for the axion.
Furthermore, 
the correlation function of the two electric 
1-form symmetry generators 
leads to the 2-form symmetry generator, 
which can be understood as the 
anomalous Hall effect of
the axionic string.
By the structure of the correlation functions, 
we have shown that the higher-form symmetries 
have a 3-group structure.
In particular, we have argued that the Witten effect and 
anomalous Hall effect  
are consequences of the action of a 0-form 
symmetry generator and the Peiffer lifting, respectively.

There are several avenues for future work.
One  is to explore a higher-group structure of the 
axion-photon system including axionic domain walls 
in the presence of the potential of the axion.
The topological nature of 
axionic domain walls can be effectively described by a 
3-form gauge theory and 3-form symmetry~\cite{Hidaka:2019mfm}.
Therefore, there would be a higher-group structure in the 
axionic domain wall.

Also, there might be 
a new interrelation between the axion electrodynamics and 
string theory.
In fact, the axion electrodynamics has been 
treated as a simple model of the string theory~\cite{Townsend:1993wy,Harvey:2000yg}.
The background gauging of 
the higher-form symmetries is an important issue,
since it can clarify 't Hooft anomalies 
as well as higher-group 
structures~\cite{Cordova:2018cvg,Benini:2018reh,Tanizaki:2019rbk}.
We will address this issue in the forthcoming paper~\cite{HNY:inprep}.

\subsection*{Acknowledgements}
R.~Y.~thanks Yuji Hirono, Taro Kimura, and Naoki Yamamoto for discussions. Y.~H. thanks Hiroki Kodama for useful comments.
This work is supported in part by Japan Society of Promotion of Science (JSPS) Grant-in-Aid for Scientific Research
(KAKENHI Grants No.~17H06462, 18H01211 (Y.~H.) and 18H01217 (M.~N)).

\end{document}